\documentclass[a4paper]{article}
\usepackage{a4wide}
\usepackage{epsfig}

\def\be{\begin{equation}}
\def\ee{\end{equation}}
\def\bea{\begin{eqnarray}}
\def\eea{\end{eqnarray}}

\title{Quantum field dynamics of the slow rollover in the
    linear delta expansion}

\author{D.~J.~Bedingham\thanks{email:
{\tt d.j.bedingham@sussex.ac.uk }}  \\
{\small\it Centre for Theoretical Physics, University of Sussex,}\\
{\small\it Falmer, Brighton BN1 9QJ, United Kingdom.} \\  \\
H.~F.~Jones\thanks{email:
{\tt H.F.Jones@imperial.ac.uk}}\\
{\small\it Physics Department, Imperial College,}\\
{\small\it  London SW7 2BZ, United Kingdom.}
}

\date{\today}

\begin{document}
\maketitle

\begin{abstract}
We show how the linear delta expansion, as applied to the slow-roll
transition in quantum mechanics,
can be recast in the closed time-path formalism. This results in
simpler, explicit expressions than were obtained in the
Schr\"odinger formulation and allows for a straightforward
generalization
to higher dimensions.
Motivated by the success of the method in the quantum-mechanical
problem, where it has been shown to give
more accurate results for longer than existing alternatives,
we apply the linear delta expansion to four-dimensional
field theory.

At small times all methods agree. At later times, the first-order
linear delta expansion is consistently higher that Hartree-Fock,
but does not show any sign of a turnover. A turnover emerges in
second-order of the method, but the value of $\langle
\hat{\Phi}^2(t) \rangle$ at the turnover is larger that that given
by the Hartree-Fock approximation. Based on this calculation, and
our experience in the corresponding quantum-mechanical problem, we
believe that the Hartree-Fock approximation does indeed
underestimate the value of $\langle \hat{\Phi}^2(t) \rangle$ at
the turnover. In subsequent applications of the method we hope to
implement the calculation in the context of an expanding universe,
following the line of earlier calculations by Boyanovsky {\sl et
al.}, who used the Hartree-Fock and large-$N$ methods. It seems
clear, however, that the method will become unreliable as the
system enters the reheating stage.

\end{abstract}

%\vspace{-6.5in} \hfill SUSX-TH-02-018 \vspace{6.2in}

\section{Introduction}
A period of inflation in the early universe could have the desirable
consequence that a general initial condition will evolve towards
the homogeneity, isotropy and flatness which we observe.
Basic models require the slow evolution of a scalar
field from an initial unstable vacuum state to a final stable state.
Without knowing how to perform this inherently non-perturbative
calculation exactly, approximation attempts must
first prove themselves in the simpler situation of the
quantum-mechanical slow roll.
Though this simpler problem cannot be solved analytically, the degrees of
freedom are sufficiently few that an exact numerical solution can
be found. This allows us to test non-perturbative methods before
proceeding to a calculation for the four-dimensional scalar field.

The quantum-mechanical slow roll was first treated
by Guth and Pi \cite{GP}, who considered the evolution of a
Gaussian wave-packet initially centred at the top of a potential hill
$V=-\frac{1}{2}m\omega q^2$. Following this, the Dirac
time-dependent variational method was used for a potential
\mbox{$V=\lambda(q^2-a^2)^2/24$}, first by Cooper et al.~\cite{COOP},
who used a Gaussian wave function ansatz, and later by
Cheetham and Copeland \cite{ED}, who included the second-order
Hermite polynomial in their ansatz.

The work presented here is based on an alternative variational
approach, the linear delta expansion (LDE), recently applied \cite{HFJ}
to the quantum mechanical slow roll. The
method was found to reproduce the exact time dependence for longer
than any of the alternative methods.

In this paper we reformulate the LDE method in terms of a path integral
rather than solving the Schr\"odinger
equation with some wavefunction ansatz.
Since we directly calculate expectation values without
calculating the wavefunction, we save on calculational effort.
More importantly, it is relatively straightforward to generalize to the
generating functional formalism of quantum field theory in four
space-time dimensions. This strategy is the same as that employed by
Boyanovsky et al. in Ref.~\cite{BOY}, who were able to generalize the
Hartree method of Ref.~\cite{COOP}. Since the LDE method is more
successful
in the quantum mechanical case, we should expect it to be more accurate
when applied to field theory.

We first consider the slow-roll phase
transition in a
one-dimensional field theory (quantum mechanics) with
potential $V=-\frac{1}{2}m\omega q^2$. This
serves as a simple introduction to the path integral formulation of
this problem.
We then turn to a potential of the form
$V=\lambda(q^2-a^2)^2/{24}$
where we outline the LDE method. Finally we  demonstrate the use
of this
method for a four-dimensional scalar field undergoing an instantaneous
temperature
quench.

\begin{figure}[t]
\epsfxsize=1.5in
\centerline{\epsfbox{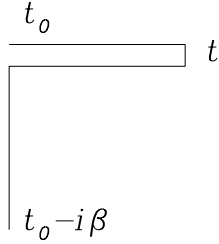}}
\caption{\small Complex time
path.} \label{fig:timepath}
\end{figure}

In line with previous papers on the quantum-mechanical slow roll,
we characterize the dynamical
process by considering the expectation value of the field operator
squared $\hat{q}^2(t)$ (now working in the Heisenberg picture)
with respect to an initial harmonic oscillator ground state. This
is equivalent to the zero-temperature limit for an initial thermal
distribution of states with Hamiltonian
$H=\frac{p^2}{2m}+\frac{1}{2}m\omega_i^2q^2$. We formulate the
problem in this way in order to facilitate our transition to
finite-temperature four-dimensional field theory. We have
    \bea
    \langle 0 \mid \hat{q}^2(t) \mid 0\rangle
    &=&\lim_{\beta\rightarrow\infty}\int{\rm d}q'
    \langle q';t_0\mid \exp\{-\beta\hat{H}\}\hat{q}^2(t)
    \mid q';t_0\rangle\\
    &=&\lim_{\beta\rightarrow\infty}\int{\rm d}q'
    \langle q';t_0-i\beta\mid \hat{q}^2(t)
    \mid q';t_0\rangle.
    \eea
Green functions with respect to an initial field state at time $t_0$
and
a final state at time $t_0-i\beta$  can be derived from a
generating functional whose time contour $c$ passes between these
two points. The contour must also pass through the time $t$ at
which the $\hat{q}^2(t)$ operator is inserted.
The time contour typically passes from $t_0$ along
the real time axis in the positive direction the point $t$ or beyond
it.
It then passes back along the real time axis to $t_0$
before moving in the imaginary time direction to $t_0-i\beta$
(see Fig.~\ref{fig:timepath}).
%%%Fig1

The generating functional is
    \be
    Z[j]=\int{\rm D}q \exp\left\{\frac{i}{\hbar}\int_c {\rm
d}t[{L}+\hbar j
    q]\right\}.
\label{eq:gener}
    \ee
The Lagrangian $L$ must satisfy
\be
L({\rm Re} \{t\}=t_0) = -\frac{1}{2}q(m\partial^2_t+m\omega_i^2)q
\ee
in order to meet the initial conditions.
At later times the form of the Lagrangian may change, modelling
some external influence on the particle.

The field boundary conditions are fixed such that
$q(t_0)=q(t_0-i\beta)$ and we derive general time contour ordered
expectation values as follows:
    \be
    \langle 0 \mid T_c\hat{q}(t_1)\hat{q}(t_2)\cdots\mid 0\rangle=
    \frac{1}{Z[0]}
    \left[\frac{\delta}{i\delta j(t_1)}
    \frac{\delta}{i\delta j(t_2)}\cdots Z[j]\right]_{j=0}
    \ee
where we take $\beta$ to infinity in the quantum-mechanical slow roll,
but in principal could choose any
value representing some fixed initial temperature. This we do when considering
the case of four-dimensional field theory.
The method is known
as the closed time path method for studying real time dependent
Green functions. It was first conceived by Schwinger \cite{SCHW}
and Keldish \cite{KELD} (for a more recent account see
\cite{BOY}).

In section 2 we outline the closed time path method and apply it
to the quantum-mechanical model. We reproduce previous results for
the inverted harmonic oscillator. In section 3 we develop the LDE
approximation method. In section 4 we apply these techniques to
four-dimensional scalar field theory in both first and second
order. For completeness we include an Appendix on the derivation
of the propagator $D(t,t')$, although this material can also be
found in standard references (\cite{LeBellac}, \cite{Landsman}).

In the remaining pages we shall use units where $\hbar=m=1$.

%%%%%%%%%%%%%%%%%%%%%%%%%%%%%%%%%%%%%%%%%%%%%%%%%%%%%%%%%%%%%%%%%%%%%%%%%%%%

\section{Inverted harmonic oscillator}
In previous articles considering the quantum mechanical slow roll
\cite{GP,COOP,ED,HFJ}, the particle begins
at a time $t=0$ when it is described by a Gaussian wave function,
centred at the top
of a potential hill. To reproduce this situation here, we consider
the particle prepared at $t<0$ in the ground state of an harmonic
oscillator potential (corresponding to a Gaussian wave function).
When $t=0$ we suddenly change the Hamiltonian to one with a
potential hill. Subsequent real time evolution of the particle
sees it ``rolling off'' the top of the hill.

First let us consider a final potential of the form
$V=-\frac{1}{2}\omega_f^2q^2$. In terms of the Lagrangian we have
    \bea
    {L}(t)&=&\frac{1}{2}q{K}(t) q\nonumber\\
    {K}(t)&=&-\partial^2_t-\omega^2(t)\nonumber\\
    \omega^2(t)&=&\Theta(-t)\omega_i^2-\Theta(t)\omega_f^2.
    \eea
where we define
    \be
    \Theta(t)=\left\{\begin{array}{ll}
    1 & {\rm Re}\{t\}>0\\
    0 & {\rm Re}\{t\}<0
    \end{array}\right. .
    \ee

To solve the field theory we begin by shifting the field variable
$q$ in Eq.~(\ref{eq:gener}) in order to complete the square
    \be
    q(t)\rightarrow q(t)-\int_c {\rm d}t' D(t,t')j(t').
    \ee
The propagator $D$ must satisfy $K(t)D(t,t')=\delta_c(t,t')$,
where the contour delta function
$\delta_c(t,t')$ is defined for a test function $f(t)$ by $
\int_c{\rm d}t'f(t')\delta_c(t,t')=f(t)$. This results in a
generating functional of the form
    \be
    Z[j]=Z[0]\exp\left\{-i\int_c{\rm d}t'{\rm d}t''\left[
    \frac{1}{2}j(t')D(t',t'')j(t'')\right]\right\}.
    \ee
Performing the functional derivatives in order to obtain
$\langle\hat{q}^2(t)\rangle$ we find
    \be
    \langle\hat{q}^2(t) \rangle=
    \langle 0 \mid \hat{q}^2(t) \mid 0\rangle=
    \frac{1}{Z[0]}
    \left[-\frac{\delta^2}{\delta j^2(t)}Z[j]\right]_{j=0}=
    iD(t,t).
    \label{eq:q2=D}
    \ee
In the zero temperature limit, the propagator is found to
have the general solution (see appendix)
    \be
    i D(t_1,t_2)=\frac{1}{2\omega_i}\left[
    \theta_c(t_1-t_2)U^{-}(t_1)U^{+}(t_2)
    +\theta_c(t_2-t_1)U^{+}(t_1)U^{-}(t_2)\right]
    \label{eq:D}
    \ee
where
    \be
    [\partial^2_t+\omega^2(t)]U^{\pm}(t)=0
    \label{eq:modeeqn}
    \ee
and for ${\rm Re}\{t\}<0$ the two independent solutions are
    \be
    U^{\pm}(t)=\exp\{\pm i\omega_i t\}.
    \ee
The dynamical information of the theory is contained purely in the
$U^{\pm}$-functions. The problem is essentially reduced to solving
a second order differential equation. Fixing the boundary conditions
such that
    \bea
    U^{\pm}(0+)&=&U^{\pm}(0-)\\
    \partial_t U^{\pm}(0+)&=&\partial_t U^{\pm}(0-)
    \eea
we find the general solution to Eq.~(\ref{eq:modeeqn}):
    \bea
    U^{\pm}(t)=\Theta(-t) e^{\pm i\omega_i t}
%   \nonumber\\&&
    +\Theta(t)\left(\cosh(\omega_f t)\pm i{\omega_i
    \over\omega_f}\sinh(\omega_f
    t)\right).
    \label{eq:U}
    \eea
Putting Eqs.~(\ref{eq:q2=D}), (\ref{eq:D}) and (\ref{eq:U})
together we find
    \be
    \langle\hat{q}^2(t) \rangle=\Theta(-t)\frac{1}{2\omega_i}+
    \Theta(t)\frac{1}{2\omega_i}\left[1+\frac{1}{2}
    \left(1+\frac{\omega_i^2}{\omega_f^2}\right)\left[\cosh(2\omega_f
t)-1\right]
    \right].
    \label{eq:iho}
    \ee

This is the standard harmonic oscillator result for $t<0$. For $t>0$
the
expectation value begins to grow as the particle rolls off the top
of the hill. The growth becomes exponential for large $t$:
    \be
    \langle\hat{q}^2(t) \rangle\rightarrow
    \frac{1}{8\omega_i}\left(1+\frac{\omega_i^2}{\omega_f^2}\right)
    \exp(2\omega_f t).
    \ee
This is in exact agreement with Guth and Pi \cite{GP} after carefully
comparing parameters.

%%%%%%%%%%%%%%%%%%%%%%%%%%%%%%%%%%%%%%%%%%%%%%%%%%%%%%%%%%%%%%%%%%%%%%%%%%%%%%

\section{Linear delta expansion}
We next turn to the problem of a symmetry breaking potential described
by a Lagrangian of the form
    \bea
    {L}(t)&=&\frac{1}{2}q{K}(t) q-\frac{\lambda(t)}{24}q^4\nonumber\\
    {K}(t)&=&-\partial^2_t-\omega^2(t)\nonumber\\
    \omega^2(t)&=&\Theta(-t)\omega_i^2-\Theta(t)\omega_f^2\nonumber\\
    \lambda(t)&=&\Theta(t)\lambda.
    \eea

We could at this stage perform a perturbative expansion in powers
of $\lambda$. However, we know that the particle is bound by the $q^4$
term
to a region near to $q=0$. If we
perturb about the Gaussian solution for $\langle\hat{q}^2(t)
\rangle$, the perturbative correction must become large so as to
prevent the exponential increase, and the philosophy of perturbation
theory therefore breaks down.

The linear delta expansion (LDE) is a practical way of improving
those aspects of a perturbative series which lead to its
divergence \cite{PERN,Yukalov}. In toy models, where exact results
are achievable, the LDE is known to produce convergent results and
to do so much faster than alternatives. See, for instance,
\cite{DJQM,DJAHO} and references therein. The LDE has also been
used successfully in many other situations, including studies of
scalar theories \cite{ME}.

In practice we substitute the Lagrangian with a new
$\delta$-Lagrangian which is the same as the original upon setting
$\delta$ equal to 1
    \be
    {L}\rightarrow {L}_{\delta}=(1-\delta){L}_0+\delta{L}.
    \ee
Here, ${\cal L}_0$ is just taken to be the quadratic part of the
Lagrangian, depending on some variational mass $\mu$,
    \be
    {L}_0=-\frac{1}{2}q(\partial^2_t-\mu^2) q.
    \ee
The mass $\mu$ is treated as a constant for the purpose of performing
any time integrals, and $\mu^2$ is taken to be equal to $-\omega_i^2$ for ${\rm Re}\{t\}<0$ so
as not to interfere
with the fixed initial conditions. We have
    \be
    {L}({\rm Re}\{t\}<0)=-\frac{1}{2}q(\partial^2_t+\omega_i^2) q
    \ee
and
    \be
    {L}_{\delta}({\rm Re}\{t\}>0)=-\frac{1}{2}q(\partial^2_t-\mu^2) q
    +\delta\left[\frac{(\omega_f^2-\mu^2)}{2}q^2
    -\frac{\lambda}{24}
    q^4 \right].
    \ee
Any given physical quantity is calculated as a perturbative
expansion up to some given order in $\delta$. We then set $\delta$
equal to $1$ and choose the value of $\mu$ according to the
principle of minimal sensitivity (PMS). For $\langle \hat{q}^2(t)
\rangle$ this is
    \be
    \frac{{\rm d}\langle \hat{q}^2(t) \rangle}{{\rm d}\mu}=0.
    \label{eq:PMS}
    \ee
The rationale for the PMS is that, although the exact value of the
quantity in question can not depend on $\mu$, the expansion will
have some residual $\mu$ dependence when truncated to some finite
order. The stationary points have a special status, in that at
such points this dependence is locally zero. At other points,
where the dependence is non-zero, there is no reason to choose one
over another. Apart from this logical justification, it has been
rigorously proved in some simple models that the sequence of
approximations\footnote{It is important to note that the LDE with
PMS gives a sequence of approximations, rather than a conventional
series, in which subsequent orders merely add additional terms to
the series. Instead, subsequent orders also change the values of
earlier terms.} provided by the PMS indeed converges
(exponentially rapidly) to the exact answer\cite{DJQM,DJAHO}, in
contrast to the perturbative expansion, where $\mu$ is fixed,
which gives rise to an alternating divergent series. Apart from
these proofs of convergence, it has been applied successfully, in
a pragmatic way, to a large variety of problems in quantum
mechanics and quantum field theory, both in the continuum and on
the lattice\cite{VCE,PJ}.

In some problems it is unfortunately the case that there is not a
unique solution to the PMS condition, i.e. that there are several
stationary points. In that event, some element of subjective
judgement has to be exercised, such as the width of the maximum or
minimum and continuity with known results or expected behaviour.

In the present problem the PMS criterion provides a different
constraint on $\mu$ for each final time that we consider. Though
$\mu$ will be different for different final times, it is not
considered as a time dependent function in the evolution up to
that final time. This is the simplest and most natural way to
implement the LDE in a time-dependent problem.

The propagator is given as in Eq.~(\ref{eq:D}); however, the mode
functions
are now dependent on $\mu$ and satisfy
    \be
\left[\partial^2_t+\Theta(-t)\omega_i^2-\Theta(t)\mu^2\right]U^{\pm}(t)=0,
    \ee
with solution
    \be
    U^{\pm}(t)=\Theta(-t) e^{\pm i\omega_i t}
    +\Theta(t)\left(\cosh(\mu t)\pm i{\omega_i \over\mu}\sinh(\mu
t)\right).
    \ee
%%%Fig2
\begin{figure}[h]
\epsfxsize=15pc
\centerline{\epsfbox{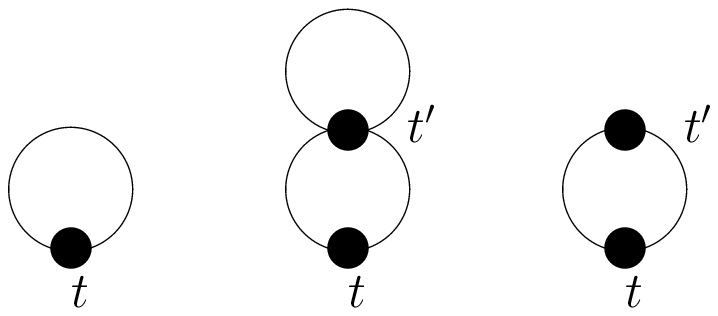}} \caption{\small
Contributions to $\langle \hat{q}^2(t) \rangle$.}
\label{fig:DIAGS}
\end{figure}
At first order in $\delta$, the relevant Feynman diagrams
(Fig.~\ref{fig:DIAGS})
can be written out to give
    \bea
    \langle \hat{q}^2(t) \rangle
    &=&iD(t,t)\nonumber\\ &&
    +\frac{\delta\lambda}{2}\int_c
    {\rm d}t'\Theta(t') iD^2(t',t)iD(t',t')
    \nonumber\\ && -\delta(\omega_f^2-\mu^2)\int_c{\rm d}t'\Theta(t')
    iD^2(t',t).
    \label{eq:qsquared}
    \eea
Evaluating the integrals we have (for $t>0$)
    \bea
    \langle \hat{q}^2(t) \rangle &=&
    \frac{1}{2\omega_i}\left[1+\frac{1}{2}
        \left(1+\frac{\omega_i^2}{\mu^2}\right)\left[\cosh(2\mu
t)-1\right]
        \right]\nonumber\\
    &&\!\!\!\!\!\!\!\!\!\!\!\!\!\!
+\frac{\lambda}{16^2\omega_i^2\mu^2}\left[8\frac{\omega_i^2}{\mu^2}
\left(1-\frac{\omega_i^2}{\mu^2}\right)
    \left[\cosh(2\mu t)-1\right]
    \right. \nonumber\\  && \left.
\;\;\;\;\;\;\;\;\;\;-12
\left(1-\frac{\omega_i^4}{\mu^4}\right)\mu t\sinh(2\mu t)
-\left(1+\frac{\omega_i^2}{\mu^2}\right)^2\left[\cosh(4\mu t)-1\right]
    \right]
    \nonumber\\ &&\!\!\!\!\!\!\!\!\!\!\!\!\!\!
    +\frac{(\omega_f^2-\mu^2)}{4\omega_i\mu^2}
    \left[\frac{\omega_i^2}{\mu^2}\left[
    1-\cosh(2\mu t)\right]+\left(1+\frac{\omega_i^2}{\mu^2}\right)\mu t
    \sinh(2\mu t)\right].
    \label{eq:answer}
    \eea
This is a remarkably simple, explicit form for $\langle \hat{q}^2(t)
\rangle$ compared
with the complicated implicit expressions given in Ref. \cite{HFJ}.
However, we have verified
that these expressions do indeed reduce to Eq.~(\ref{eq:answer}).

\begin{figure}[h]
\epsfxsize=25pc \centerline{\epsfbox{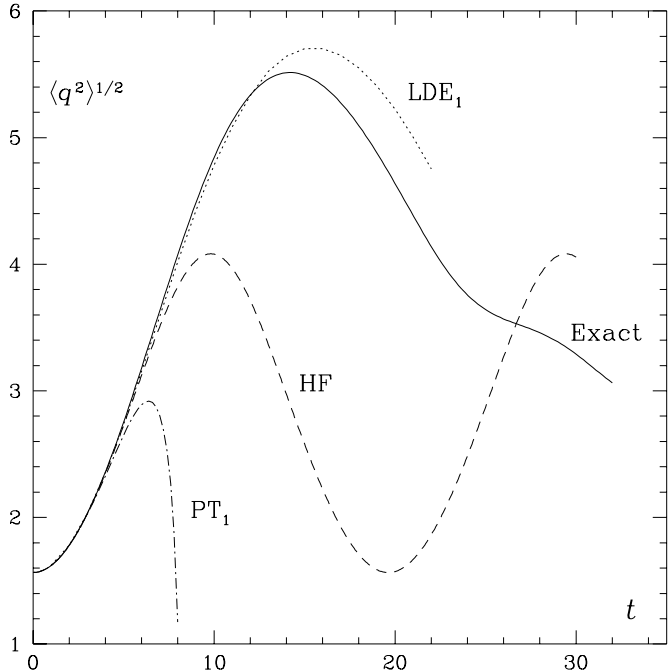}} \caption{\small
Slow roll in quantum mechanics: $\langle \hat{q}^2(t)
\rangle^{1/2}$ versus $t$. The first-order LDE result is compared
against the exact result. Also shown are the Hartree-Fock results
of Ref.~\cite{COOP} (HF), and first-order perturbation theory
(${\rm PT}_1$).} \label{fig:QM}
\end{figure}

To proceed, we find the optimum value of Eq.~(\ref{eq:answer})
according to
the PMS criterion, Eq.~(\ref{eq:PMS}). The result is the curve shown in
Fig.~\ref{fig:QM}.
We have chosen $\lambda=0.01$ and $w_i^2=w_f^2={25\lambda}/{6}$
(recall that
$w_f^2$ appears with a different sign in the Lagrangian). These
parameters coincide with those chosen in \cite{GP,COOP,ED,HFJ} in
order that we may easily compare our results.
Also shown are the exact result, first-order perturbation
theory, and the Hartree approximation of
Ref.~\cite{COOP}.

First-order perturbation theory is achieved
upon setting $\mu^2=\omega_f^2$ in Eq.~(\ref{eq:answer}), while the
Hartree approximation amounts to taking
$\mu$ to be a time-dependent function given by
$\mu^2(t')=\omega^2_f-({\lambda}/{2})i D(t',t')$. This results in
a cancellation between the coupling correction and the mass insertion,
and a self-consistent set of equations
    \bea
        \langle \hat{q}^2(t) \rangle = iD(t,t) =
    \frac{1}{2\omega_i}U^{-}(t)U^{+}(t)\nonumber \\
    \left[\partial^2_t+\omega^2(t)+\frac{\lambda}{2} iD(t,t)\right]
    U^{\pm}(t)=0.
    \eea

%%%%%%Fig3

The LDE result is seen to track the exact result for a
significantly longer time than the Hartree result. It then
overshoots, signifying that the LDE result gives a much improved
description of the inflationary period, but does not do so well
during reheating.

In quantum mechanics it is possible to go to high order in the LDE
by the use of recursion relations. The results of this exercise
were given in Ref.~\cite{HFJ}, where the calculations were carried
out to O($\delta^7$). It turns out that the second- and
third-order calculations do not exhibit clear PMS points, but
thereafter successive orders follow the true curve more and more
accurately up to the turnover point, but diverge beyond that
point. We can hope that in field theory in (3+1) dimensions, the
LDE will again give a good description of the initial slow-roll
process. In field theory, however, it is not practical to go
beyond second order.

%***
%%%%%%%%%%%%%%%%%%%%%%%%%%%%%%%%%%%%%%%%%%%%%%%%%%%%%%%%%%%%%%%%%%%%%%%%%
%%%%%%%%%%%%%%%%%%%%%%%%%%%%%%%%%%%%%%%%%%%%%%%%%%%%%%%%%%%%%%%%%%%%%%%%%

\section{Scalar field theory}

Having developed our method for quantum mechanics, it remains to
see how easily it can be implemented for the case of field theory.
We consider a single real scalar field theory with time-dependent
Lagrangian of the form
    \bea
    L(t)&=&\int{\rm
d}^3x\left\{\frac{1}{2}\Phi\left(-\partial^2_t+\nabla^2-m^2(t)\right)\Phi
    -\frac{\lambda}{24}\Phi^4\right\}\\
    m^2(t)&=&\Theta(-t)m_i^2-\Theta(t)m_f^2.
    \eea
With appropriate choice of the parameters, this model crudely
describes a sudden temperature quench in which the field is driven
through a phase transition at time $t=0$.

Our interest is in determining the quantity
    \be
    \langle \hat{\Phi}^2(t)\rangle =\frac{1}{V}\int{\rm d}^3 x
    \langle\hat{\Phi}^2({\bf x},t)\rangle.
    \ee
To perform the delta expansion we again define a
$\delta$-Lagrangian by
    \be
    {L}\rightarrow {L}_{\delta}=(1-\delta){L}_0+\delta{L}
    \ee
with
    \be
    \label{L0}
    {L}_0=\int{\rm d}^3 x \;\; \left\{\frac{1}{2}\Phi
    \left(-\partial^2_t+\nabla^2+\mu^2\right)\Phi\right\}.
    \ee
We replace the original Lagrangian by our $\delta$-Lagrangian for
${\rm Re}\{t\}>0$. This gives
    \bea
    {L}({\rm Re}\{t\}>0)&=&\int{\rm d}^3 x \left\{
\frac{1}{2}\Phi{\left(-\partial^2_t+\nabla^2+\mu^2\right)}\Phi\right.\nonumber\\
&&\left.+\delta\left[\frac{(m^2_f-\mu^2)}{2}\Phi^2-\frac{\lambda}{24}
    \Phi^4\right]\right\}.
    \eea

Switching to momentum space, the propagator now satisfies the
relation
    \be
    {K}_{\rm p}(t)D_{\rm p}(t,t')=\delta_c(t,t'),
    \ee
where
    \bea
    {K}_{\rm p}(t)&=&-\partial^2_t-\omega_{\rm p}^2(t)\\
    \omega_{\rm p}^2(t)&=& \Theta(-t)\omega_{i;{\rm
p}}^2-\Theta(t)\omega_{f;{\rm p}}^2,
    \eea
and now
    \bea
    \omega_{i;{\rm p}}^2=p^2+m_i^2\\
    \omega_{f;{\rm p}}^2=\mu^2-p^2.
    \eea
The propagator has the solution (see appendix)
    \be
    i D_{\rm p}(t_1,t_2)=\theta_c(t_1-t_2)iD_{\rm p}^>(t_1,t_2)
    +\theta_c(t_2-t_1)iD_{\rm p}^<(t_1,t_2)
    \ee
where
    \bea
    iD_{\rm
    p}^>(t_1,t_2)=\frac{1}{2\omega_{i;{\rm
p}}}\frac{1}{e^{\omega_{i;{\rm p}}
    \beta}-1}\left[U^+_{\rm p}(t_1)U^-_{\rm
    p}(t_2)+e^{\omega_{i;{\rm p}}\beta}U^-_{\rm p}(t_1)U^+_{\rm
    p}(t_2)\right]\\
    iD_{\rm
    p}^<(t_1,t_2)=\frac{1}{2\omega_{i;{\rm
p}}}\frac{1}{e^{\omega_{i;{\rm p}}
    \beta}-1}\left[e^{\omega_{i;{\rm p}}\beta}U^+_{\rm p}(t_1)U^-_{\rm
    p}(t_2)+U^-_{\rm p}(t_1)U^+_{\rm
    p}(t_2)\right].
    \eea
The mode functions satisfy
    \be
    \left[\partial^2_t+\omega_{\rm p}^2(t)\right]U_{\rm p}^{\pm}(t)=0,
    \ee
with solutions
    \be
    U_{\rm p}^{\pm}(t)=\Theta(-t)e^{\pm i\omega_{i;{\rm p}} t}
    +\Theta(t)\left(\cosh(\omega_{f;{\rm p}} t)\pm i{\omega_{i;{\rm p}}
\over\omega_{f;{\rm p}}}
    \sinh(\omega_{f;{\rm p}} t)\right).
    \ee
\subsection{First order}
The same diagrams which contributed to $\langle \hat{q}^2(t)
\rangle$ in the previous section contribute to $\langle
\hat{\Phi}^2 (t) \rangle$ here.
The essential difference from the quantum-mechanical case is that
the propagators now depend on momentum and that any loops will
involve an integration over loop momenta. The Feynman diagrams in
Fig.~\ref{fig:DIAGS} give
    \bea
    \langle \hat{\Phi}^2(t) \rangle
    &=&\int_{\rm p}iD_{\rm p}(t,t)\nonumber\\
    &&+\frac{\delta\lambda}{2}\int_c{\rm
    d}t' \int_{\rm p}iD_{\rm p}^2(t',t)\int_{\rm k}iD_{\rm k}(t',t')
    \nonumber\\ &&- \delta(m_f^2-\mu^2)\int_c{\rm
    d}t' \Theta(t')\int_{\rm p} iD_{\rm p}^2(t',t)
    \eea
(cf. Eq.~(\ref{eq:qsquared})) where we have used the notation
    \be
    \int_{\rm p}=\int\frac{{\rm d}^3 p}{(2\pi)^3}.
    \ee

The momentum integrals are
divergent and must be regularized. As in Ref.~\cite{BOY}, we
assume a scheme which leaves the contributions from stable modes
($p^2,k^2>\mu^2$) being negligibly small. The dominant growth in
$ \langle \hat{\Phi}^2(t)\rangle$ is associated with the finite
contribution of the unstable modes. In practice this means that we
may perform momentum integrals in the finite range $p^2,k^2<\mu^2$
to achieve finite results. We simply make the replacement
    \be
    \int_{\rm p}=\frac{1}{2\pi^2}\int_0^{\mu} p^2{\rm d}p.
    \ee
The calculations are performed
in the high-temperature limit, where $\beta\omega_{i;{\rm p}}\ll 1$, so that
    \be
    \coth \left( \frac{1}{2} \omega_{i;{\rm p}} \beta
\right)\sim\frac{2}
    {\beta \omega_{i;{\rm p}} }.
    \ee
In this limit the first term is found to be
    \be
    \frac{1}{\beta}\int_{\rm p}\frac{1}{\omega_{i;{\rm p}}^2}
    \left[1+\frac{1}{2}
        \left(1+\frac{\omega_{i;{\rm p}}^2}
    {\omega_{f;{\rm p}}^2}\right)\left[\cosh(2\omega_{f;{\rm p}}
t)-1\right]
     \right],
    \ee
the second is
    \bea
    &&\!\!\!\!-\frac{\lambda}{2\beta^2}\int_0^t{\rm d}t'
    \int_{\rm p}\frac{1}{\omega_{i;{\rm p}}^2\omega_{f;{\rm p}}}
    \sinh\left[\omega_{f;{\rm p}}(t-t')\right]
    \nonumber\\ &&\times\left[\left(
    1+\frac{\omega_{i;{\rm p}}^2}{\omega_{f;{\rm p}}^2}\right)
    \cosh\left[\omega_{f;{\rm p}}(t+t')\right]+\left(
    1-\frac{\omega_{i;{\rm p}}^2}{\omega_{f;{\rm p}}^2}\right)
    \cosh\left[\omega_{f;{\rm p}}(t-t')\right]\right]
    \nonumber\\ &&\times \int_{\rm k}\frac{1}{\omega_{i;{\rm k}}^2}
    \left[1+\frac{1}{2}
        \left(1+\frac{\omega_{i;{\rm k}}^2}{\omega_{f;{\rm k}}^2}
    \right)\left[\cosh(2\omega_{f;{\rm k}} t')-1\right]
     \right],
    \eea
and the third is
    \bea
    \frac{(m_f^2-\mu^2)}{2\beta}\int_{\rm p}\frac{1}{\omega_{i;{\rm
p}}^2
    \omega_{f;{\rm p}}^2}
    \left[\frac{\omega_{i;{\rm p}}^2}{\omega_{f;{\rm p}}^2}
    \left[1-\cosh(2\omega_{f;{\rm p}}t)\right]\right.
\quad\quad\quad\quad\quad\quad
    \nonumber\\ \left.
    +\left(1+\frac{\omega_{i;{\rm p}}^2}{\omega_{f;{\rm p}}^2}
    \right)\omega_{f;{\rm p}} t\sinh(2\omega_{f;{\rm p}}t)
    \right].
    \eea
In the second term, the time integral has not been performed explicitly
since the
result is rather involved.

Finally we impose the PMS constraint at each $t$ in order to find
$\mu$ and evaluate $\langle \hat{\Phi}^2(t) \rangle$
    \be
    \frac{{\rm d}\langle \hat{\Phi}^2(t) \rangle}{{\rm d}\mu}=0.
    \ee

For numerical calculations
the units are chosen such that $\hbar=c=k_B=m_i^2=1$. The remaining
parameters
are then chosen in these units to be $m_f^2=1$,
$T={1}/{\beta}=4\surd({6}/{\lambda})$ (the initial
temperature) and $\lambda=10^{-12}$. These are chosen to coincide
with those in Ref.~\cite{BOY}. The initial temperature has no
particular meaning, it is simply twice the critical temperature.
The coupling must be small for this type of model of inflation due
to constraints from the spectrum of density fluctuations.

\begin{figure}[h!]
\epsfxsize=23pc \centerline{\epsfbox{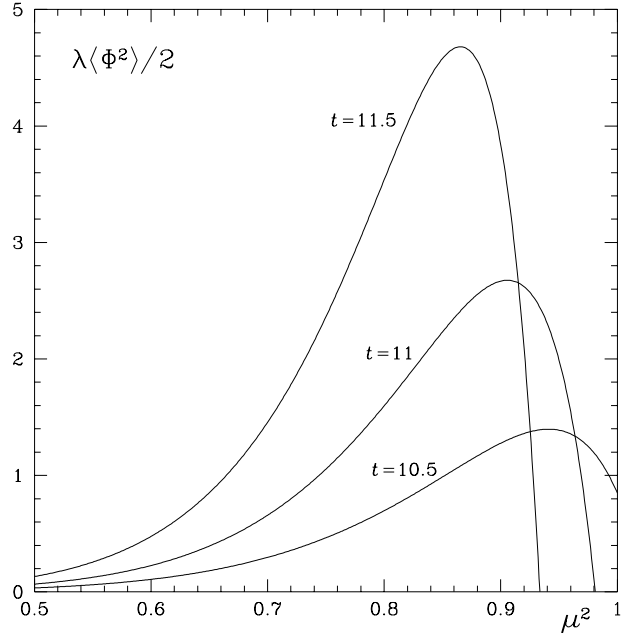}} \caption{\small
$\lambda\langle \hat{\Phi}^2(t)\rangle/2$ versus $\mu^2$ for
$t=10.5,11,11.5$ in first-order LDE. } \label{PMS1}
\end{figure}

Examples of $\lambda\langle \hat{\Phi}^2 \rangle/2$ as functions
of $\mu^2$ for various times are shown in Fig.~\ref{PMS1}. We
observe a single stationary point, a maximum, which moves to the
left and becomes sharper as $t$ increases. The motivation for the
PMS criterion is that the exact answer is independent of $\mu$. In
any finite order of the LDE this independence can only be achieved
locally.  A broad maximum indicates that the LDE is robust, but it
becomes increasingly unreliable as the peak becomes sharper. From
Fig.~\ref{PMS1} we estimate that the first-order LDE can not be
trusted beyond about $t$=11.
%%%%Fig4

\begin{figure}[h!]
\epsfxsize=22pc \centerline{\epsfbox{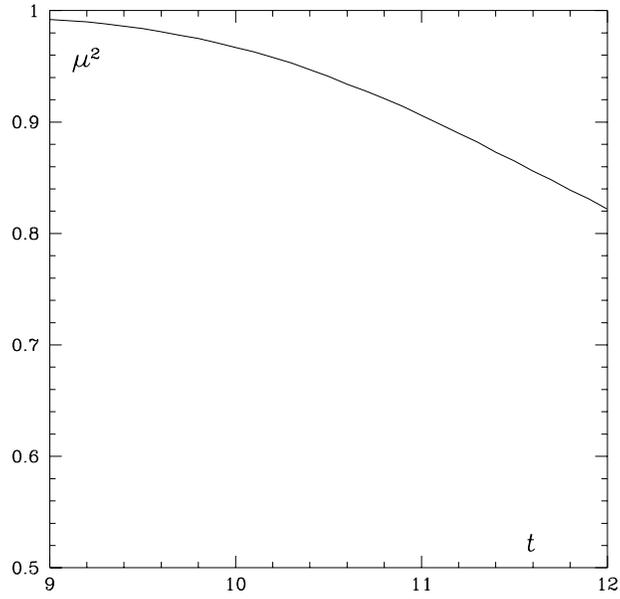}} \caption{\small
The PMS maximum $\mu^2$ versus $t$ in first-order LDE.}
\label{mu1max}
\end{figure}

The position of the maximum versus time is shown in
Fig.~\ref{mu1max}. At small times the dominant part of the action
is the quadratic part, and the evolution is well described by
perturbation theory, i.e. $\mu^2\sim m_f^2=1$. At later times, as
the fluctuations of the field grow, the quartic terms become more
important. In the context of the LDE this is taken into account by
smaller values of $\mu^2$ in the trial Lagrangian $L_0$ of
Eq.~(\ref{L0}).
%%%%Fig5

\begin{figure}[h!]
\epsfxsize=22pc \centerline{\epsfbox{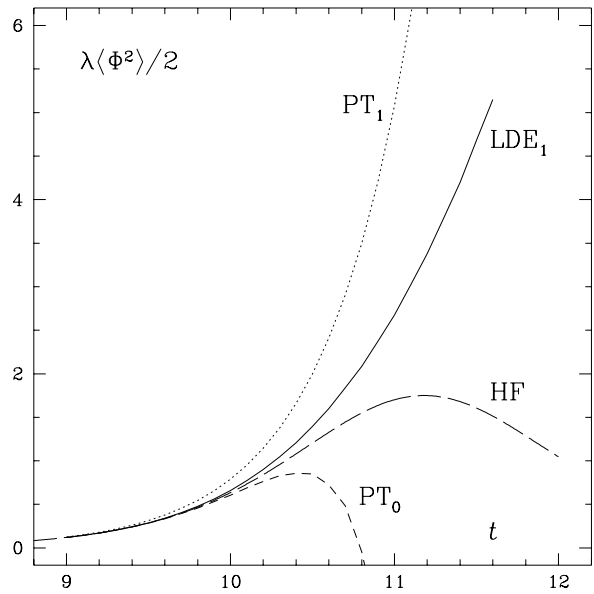}} \caption{\small
$\lambda\langle \hat{\Phi}^2(t)\rangle/2$ versus $t$. The
first-order LDE result is shown as a solid line. Also shown are
the Hartree-Fock result of Ref.~\cite{BOY} (HF) and zeroth (${\rm
PT}_0$) and first-order (${\rm PT}_1$) perturbation theory.}
\label{fig:plot1}
\end{figure}

The results for the evolution of the field are shown in
Fig.~\ref{fig:plot1}, in the restricted range of $t$ where the
different methods begin to diverge. Though we have no exact
solution to compare with, the results display the same qualitative
behaviour as in the quantum mechanical case studied in the earlier
sections (Fig.~\ref{fig:QM}).

First-order perturbation theory is achieved within the LDE
framework by setting $\mu^2=m_f^2$.
As in the quantum-mechanical case, the
Hartree
result can be reproduced by considering $\mu$ to be a time-dependent
function, this time
given by
$$
\mu^2(t')=m^2_f-({\lambda}/{2})\int_{\rm p} i D(t',t').
$$
The resulting
self-consistent set of equations are
    \bea
        \langle \hat{\Phi}^2(t) \rangle = \int_{\rm p} iD(t,t) =
    {1\over \beta}\int_{\rm p}\frac{1}{\omega_{i;{\rm p}}^2} U_{\rm
p}^{-}(t)U_{\rm p}^{+}(t)\nonumber \\
    \left[\partial^2_t+\omega_{\rm p}^2(t)+\frac{\lambda}{2} \int_{\rm
p} iD(t,t)\right]
    U_{\rm p}^{\pm}(t)=0.
    \eea
%%%Fig6

All methods give almost indistinguishable results up to $t \sim
9$. The Hartree and LDE methods remain close up to the classical
spinodal region (where $V''(\Phi)<0$, i.e.
$\frac{\lambda}{2}\langle \hat{\Phi}^2(t) \rangle>1$ ). At later
times the LDE method gives a larger value of $\langle
\hat{\Phi}^2(t) \rangle$ than the Hartree method. Based on our
experience of the quantum-mechanical case, we believe that the
Hartree method turns over prematurely and that the LDE is closer
to the exact result for longer. However, to this order it fails to
give any indication of a turnover. As mentioned in relation to
Fig.~\ref{PMS1}, the LDE becomes unreliable beyond $t\sim 11$, as
the PMS peak becomes narrower.

\subsection{Second order}

To second order in the LDE there are altogether six additional graphs. These
are exhibited in Eqs.~(\ref{newdiags4}--\ref{newdiags9}), along
with their analytic expressions, where the integrals along the
time contour have not yet been performed. We have used a more
compact notation for the $D$'s, whereby $D_{\rm p}^{t't}$ stands
for $D_{\rm p}(t',t)$ and so on. \setlength{\unitlength}{2mm}
\bea\label{newdiags4}
\begin{picture}(4,5)
        \put(2,1){\circle{2}}
        \put(2,3){\circle{2}}
        \put(2,5){\circle{2}}
        \put(2,0){\circle*{0.7}}
        \put(2,2){\circle*{0.7}}
        \put(2,4){\circle*{0.7}}
        \end{picture}
&=& \frac{-\delta^2\lambda^2}{4} \int_c {\rm d}t' \int_c {\rm
d}t''
        \int_{\rm p}\left(iD_{\rm p}^{t't}\right)^2
        \int_{\rm k_1} \left(iD_{\rm k_1}^{t''t'}\right)^2
        \int_{\rm k_2} iD_{\rm k_2}^{t''t''}\\
\begin{picture}(4,5)
        \put(2,1){\circle{2}}
        \put(3.2,2.2){\circle{1.5}}
        \put(0.8,2.2){\circle{1.5}}
        \put(2,0){\circle*{0.7}}
        \put(1.2,1.7){\circle*{0.7}}
        \put(2.8,1.7){\circle*{0.7}}
        \end{picture}
&=& \frac{-\delta^2\lambda^2}{4}\int_c {\rm d}t' \int_c {\rm d}t''
        \int_{\rm p} iD_{\rm p}^{t't}iD_{\rm p}^{t''t}iD_{\rm p}^{t't''}
        \int_{\rm k_1} iD_{\rm k_1}^{t't'}\int_{\rm k_2}iD_{\rm k_2}^{t''t''}\\
\begin{picture}(4,5)
        \put(2,1){\circle{2}}
        \put(2,2){\circle{1.8}}
        \put(2,0){\circle*{0.7}}
        \put(1.2,1.7){\circle*{0.7}}
        \put(2.8,1.7){\circle*{0.7}}
        \end{picture}
&=& \frac{-\delta^2\lambda^2}{6}\int_c {\rm d}t' \int_c {\rm d}t''
        \int_{\rm p}iD_{\rm p}^{t't}iD_{\rm p}^{t''t}
        \int_{\rm k_1,k_2}
        iD_{\rm k_1}^{t't''}iD_{\rm k_2}^{t't''}iD_{\rm p-k_1-k_2}^{t't''}\label{sunset1}\\
\begin{picture}(4,5)
        \put(2,1){\circle{2}}
        \put(2,3){\circle{2}}
        \put(2,0){\circle*{0.7}}
        \put(2,2){\circle*{0.7}}
        \put(2,4){\circle*{0.7}}
        \end{picture}
&=& \frac{\delta^2\lambda(m_f^2-\mu^2)}{2} \int_c {\rm d}t' \int_c
{\rm d}t''
        \int_{\rm p} \left(iD_{\rm p}^{t't}\right)^2 \int_{\rm k}
        \left(iD_{\rm k}^{t''t'}\right)^2\\
\begin{picture}(4,5)
        \put(2,1){\circle{2}}
        \put(0.8,2.2){\circle{1.5}}
        \put(2,0){\circle*{0.7}}
        \put(1.2,1.7){\circle*{0.7}}
        \put(2.8,1.7){\circle*{0.7}}
        \end{picture}
&=& \delta^2\lambda(m_f^2-\mu^2)
        \int_c {\rm d}t' \int_c {\rm d}t''
        \int_{\rm p} iD_{\rm p}^{t't}iD_{\rm p}^{t''t}iD_{\rm p}^{t't''}
        \int_{\rm k}iD_{\rm k}^{t't'}\\
\begin{picture}(4,5)
        \put(2,1){\circle{2}}
        \put(2,0){\circle*{0.7}}
        \put(1.2,1.7){\circle*{0.7}}
        \put(2.8,1.7){\circle*{0.7}}
        \end{picture}
&=& -\delta^2(m_f^2-\mu^2)^2
        \int_c {\rm d}t' \int_c {\rm d}t''
        \int_{\rm p} iD_{\rm p}^{t't}iD_{\rm p}^{t''t}iD_{\rm p}^{t't''}
\label{newdiags9} \eea

 In performing the time integrals in $t''$,
$t'$ over the contour of Fig.~1, the result is most easily
expressed in terms of the real and imaginary parts of the $D$'s,
or more precisely $F$ and $\rho$, defined by
\bea
F&:=&{1\over 2}\left(iD^>+iD^<\right)\nonumber\\
\rho&:=&i\left(iD^>-iD^<\right)
\eea
In the high-temperature limit, in which we are working, the imaginary
parts are much smaller than the real parts:
\bea
F_{\rm p}^{t_1t_2} &=&\frac{1}{\omega^2_{i;{\rm p}}\beta}\left[
\cosh(\omega_{f;{\rm p}} t_1)\cosh(\omega_{f;{\rm p}} t_2)
+\frac{\omega^2_{i;{\rm p}}}{\omega^2_{f;{\rm p}}}
\sinh(\omega_{f;{\rm p}} t_1)\sinh(\omega_{f;{\rm p}} t_2)\right],
\eea
compared with
\bea
\rho_{\rm p}^{t_1t_2} &=&\frac{1}{\omega_{f;{\rm p}}}\left(
\sinh(\omega_{f;{\rm p}} t_1)\cosh(\omega_{f;{\rm p}} t_2)-
\cosh(\omega_{f;{\rm p}} t_1)\sinh(\omega_{f;{\rm p}} t_2)\right).
\eea
The resulting expressions for diagrams (\ref{newdiags4})--(\ref{newdiags9}), having set $\delta=1$,
are:
\bea
        \begin{picture}(4,5)
        \put(2,1){\circle{2}}
        \put(2,3){\circle{2}}
        \put(2,5){\circle{2}}
        \put(2,0){\circle*{0.7}}
        \put(2,2){\circle*{0.7}}
        \put(2,4){\circle*{0.7}}
        \end{picture}
&=&
\lambda^2 \int_0^t {\rm d}t' \int_0^{t'} {\rm d}t''
        \int_{\rm p}F_{\rm p}^{t't}\rho_{\rm p}^{t't}
        \int_{\rm k_1} F_{\rm k_1}^{t''t'}\rho_{\rm k_1}^{t''t'}
        \int_{\rm k_2} F_{\rm k_2}^{t''t''}\\
        \begin{picture}(4,5)
        \put(2,1){\circle{2}}
        \put(3.2,2.2){\circle{1.5}}
        \put(0.8,2.2){\circle{1.5}}
        \put(2,0){\circle*{0.7}}
        \put(1.2,1.7){\circle*{0.7}}
        \put(2.8,1.7){\circle*{0.7}}
        \end{picture}
&=&
\frac{-\lambda^2}{4}\left\{ 2\int_0^t {\rm d}t' \int_0^{t'} {\rm d}t''
        \int_{\rm p} \rho_{\rm p}^{t't}F_{\rm p}^{t''t}\rho_{\rm p}^{t't''}
        \int_{\rm k_1} F_{\rm k_1}^{t't'}\int_{\rm k_2}F_{\rm k_2}^{t''t''}
\right. \nonumber\\ &&\qquad\qquad\qquad\qquad\left.
        -\int_0^t {\rm d}t' \int_0^{t} {\rm d}t''
        \int_{\rm p}\rho_{\rm p}^{t't}\rho_{\rm p}^{t''t}F_{\rm p}^{t't''}
        \int_{\rm k_1}  F_{\rm k_1}^{t't'}\int_{\rm k_2}F_{\rm k_2}^{t''t''}
        \right\}\\
        \begin{picture}(4,5)
        \put(2,1){\circle{2}}
        \put(2,2){\circle{1.8}}
        \put(2,0){\circle*{0.7}}
        \put(1.2,1.7){\circle*{0.7}}
        \put(2.8,1.7){\circle*{0.7}}
        \end{picture}
&=&
\frac{-\lambda^2}{24}\left\{ 2\int_0^t {\rm d}t' \int_0^{t'} {\rm d}t''
        \int_{\rm p}\rho_{\rm p}^{t't}F_{\rm p}^{t''t}
        \int_{\rm k_1,k_2}\left(
        12F_{\rm k_1}^{t't''}F_{\rm k_2}^{t't''}\rho_{\rm p-k_1-k_2}^{t't''}
        -\rho_{\rm k_1}^{t't''}\rho_{\rm k_2}^{t't''}\rho_{\rm p-k_1-k_2}^{t't''}
        \right)
\right.\nonumber\\ &&\left.
        +\int_0^t {\rm d}t' \int_0^{t} {\rm d}t''
        \int_{\rm p}\rho_{\rm p}^{t't}\rho_{\rm p}^{t''t}
        \int_{\rm k_1,k_2}\left(
        3F_{\rm k_1}^{t't''}\rho_{\rm k_2}^{t't''}\rho_{\rm p-k_1-k_2}^{t't''}
        -4F_{\rm k_1}^{t't''}F_{\rm k_2}^{t't''}F_{\rm p-k_1-k_2}^{t't''}\right)
        \right\}\label{sunset2}\\
        \begin{picture}(4,5)
        \put(2,1){\circle{2}}
        \put(2,3){\circle{2}}
        \put(2,0){\circle*{0.7}}
        \put(2,2){\circle*{0.7}}
        \put(2,4){\circle*{0.7}}
        \end{picture}
&=&
-2\lambda(m_f^2-\mu^2) \int_0^t {\rm d}t' \int_0^{t'} {\rm d}t''
        \int_{\rm p} F_{\rm p}^{t't}\rho_{\rm p}^{t't}\int_{\rm k}
         F_{\rm k}^{t''t'}\rho_{\rm k}^{t''t'}\\
        \begin{picture}(4,5)
        \put(2,1){\circle{2}}
        \put(0.8,2.2){\circle{1.5}}
        \put(2,0){\circle*{0.7}}
        \put(1.2,1.7){\circle*{0.7}}
        \put(2.8,1.7){\circle*{0.7}}
        \end{picture}
&=&
\frac{\lambda(m_f^2-\mu^2)}{2}
\left\{ 2\int_0^t {\rm d}t' \int_0^{t'} {\rm d}t''
        \int_{\rm p} \rho_{\rm p}^{t't}F_{\rm p}^{t''t}\rho_{\rm p}^{t't''}
        \int_{\rm k}\left( F_{\rm k}^{t't'}+F_{\rm k}^{t''t''}\right)
\right. \nonumber\\ &&\qquad\qquad\qquad\qquad\qquad\qquad\qquad\left.
        -2\int_0^t {\rm d}t' \int_0^{t} {\rm d}t''
        \int_{\rm p}\rho_{\rm p}^{t't}\rho_{\rm p}^{t''t}F_{\rm p}^{t't''}
        \int_{\rm k} F_{\rm k}^{t't'}
        \right\}\\
        \begin{picture}(4,5)
        \put(2,1){\circle{2}}
        \put(2,0){\circle*{0.7}}
        \put(1.2,1.7){\circle*{0.7}}
        \put(2.8,1.7){\circle*{0.7}}
        \end{picture}
&=&
-(m_f^2-\mu^2)^2
\left\{ 2\int_0^t {\rm d}t' \int_0^{t'} {\rm d}t''
        \int_{\rm p} \rho_{\rm p}^{t't}F_{\rm p}^{t''t}\rho_{\rm p}^{t't''}
        -\int_0^t {\rm d}t' \int_0^{t} {\rm d}t''
        \int_{\rm p} \rho_{\rm p}^{t't}\rho_{\rm p}^{t''t}F_{\rm p}^{t't''}
        \right\}
\eea
In most of the diagrams there are two terms involving an integral over $t''$
up to $t$ and another up to $t'$. In the original forms of these expressions
there were severe cancellations between the two integrals, which made accurate
integration extremely difficult. In the present, equivalent, form the two integrals give
roughly comparable contributions, posing no difficulty for numerical integration.

We have evaluated all the multidimensional integrals numerically,
including the time integrals, using the NAG Fortran routine
D01FCF. The most difficult diagram to evaluate is, of course, the
``sunset" diagram of Eqs.~(\ref{sunset1}) and (\ref{sunset2}).
Because the integrand depends only on the magnitudes of the
various momenta, there are two azimuthal integrations which can be
trivially performed, leaving a seven-dimensional integral.

The result of these calculations is that the expectation value
$\lambda\langle \hat{\Phi}^2 \rangle/2$ now develops a PMS {\sl
minimum} as a function of $\mu^2$. Examples of this behaviour are
given in Fig.~{\ref{PMS2} for the same times as were previously
shown at first order. The maximum appears to be a spurious
stationary point, with a runaway behaviour for $\lambda\langle
\hat{\Phi}^2 \rangle/2$.
%%%Fig7
\begin{figure}[h!]
\epsfxsize=20pc \centerline{\epsfbox{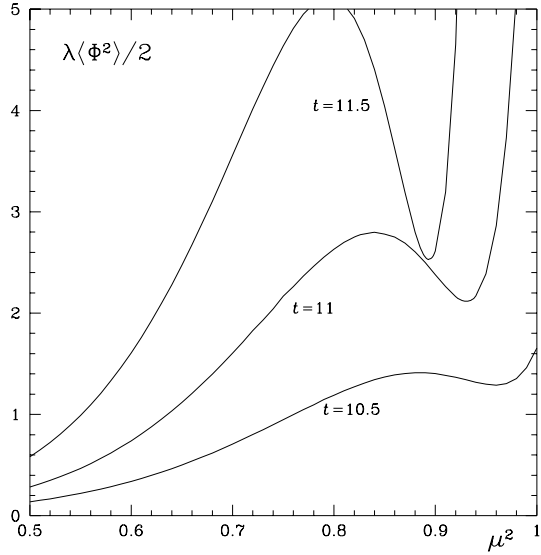}} \caption{\small
$\lambda\langle \hat{\Phi}^2(t)\rangle/2$ versus $\mu^2$ for
$t=10.5,11,11.5$ in second-order LDE. } \label{PMS2}
\end{figure}

The trend of the minimum as a function of $t$ is similar to that
of the first-order maximum, decreasing slowly as $t$ increases, as
shown in Fig.~\ref{mu2min}. The resulting plot of
$\lambda\langle\hat{\Phi}^2(t)\rangle/2$ versus $t$ is shown in
Figure \ref{LDE2}, where in addition to the Hartree-Fock result we
also show the result of the first-order large-$N$ calculation (with $N=1$).
We see that the second-order result now shows a turnover, but at a
larger value of $\langle\hat{\Phi}^2(t)\rangle$ than that given by
Hartree-Fock. This is the same feature that occurred in the
quantum-mechanical problem, and we believe gives strong evidence
that the Hartree-Fock method turns over too soon in
$\langle\hat{\Phi}^2(t)\rangle$. In this case, where there is no
symmetry breaking, the large-$N$ calculation differs from
Hartree-Fock only in that the coefficient of $iD(t,t)$ in Eq.~(54)
is reduced by a factor of 3. This means that this term takes
longer to become important and produce a turnover, so that the
maximum value is considerably greater. The same feature occurs in
the quantum mechanical problem, where the large-$N$ approximation
greatly overestimates the maximum value of
$\langle\hat{q}^2(t)\rangle$.

%%%Fig8
\begin{figure}[h!]
\epsfxsize=20pc \centerline{\epsfbox{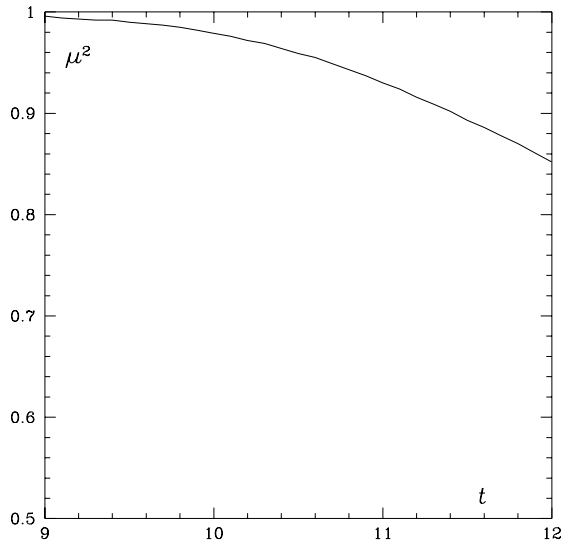}} \caption{\small
The PMS minimum $\mu^2$ versus $t$ in second-order LDE.}
\label{mu2min}
\end{figure}

%%%Fig9
\begin{figure}[h!]
\epsfxsize=20pc \centerline{\epsfbox{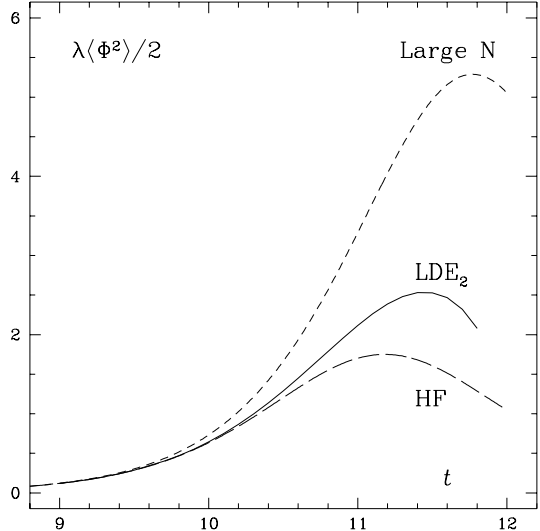}} \caption{\small
$\lambda\langle \hat{\Phi}^2(t)\rangle/2$ versus $t$. The
second-order LDE result is shown as a solid line. Also shown are
the results of Ref.~\cite{BOY} (HF), and large $N$.} \label{LDE2}
\end{figure}

We have seen that inclusion of the second-order diagrams leads to
a turnover which does not occur in first order. It would be tempting
to ascribe this turnover to the influence of the ``sunset diagram",
the first diagram to include the important effects of rescattering\cite{Berges}.
However, for the present calculation it is not possible to single out
this particular diagram from the others. Its distinctive role is rather
to provide for dissipation
and thermalization at later times (see e.g. \cite{gleiser,Boya,berges2}),
where unfortunately the LDE is unreliable. The Hartree-Fock method, which
in our language corresponds to a time-dependent $\mu$ with a particular
selection criterion, provides an example where a turnover is achieved without
the inclusion of this diagram.

%%%%%%%%%%%%%%%%%%%%%%%%%%%%%%%%%%%%%%%%%%%%%%%%%%%%%%%%%%%%%%%%%%%%%%%%
%%%%%%%%%%%%%%%%%%%%%%%%%%%%%%%%%%%%%%%%%%%%%%%%%%%%%%%%%%%%%%%%%%%%%%%%

\section{Discussion}
The main motivation for this work was to expand upon the available
machinery for tackling out-of-equilibrium problems in field
theory.

The linear delta expansion, applied to the quantum-mechanical
equivalent of the slow-roll transition, has been shown to give a
consistent improvement on other methods. However, the
Schr\"odinger formulation of Ref.~\cite{HFJ} can not immediately
be generalized to field theory in higher dimensions. We have shown
how to recast the problem in terms of the closed time-path
formalism, which can be so generalized.
This is an extension which has not been achieved in other treatments
of the quantum-mechanical problem, with the exception of the
Hartree method.

As noted in \cite{BOY},  the Hartree approximation cannot probe
the non-linear regions of the potential. Moreover, the Hartree
method is a one-off approximation, which is not capable of
systematic improvement. To understand the later time behaviour and
to probe the true vacuum, calculations must to go beyond Hartree.
The LDE, a systematic expansion with a variational component,
offers just this possibility, although for practical reasons, it
would be extremely difficult to go beyond second order in quantum
field theory.

The main result of the paper is the formalism outlined in section
4, and Fig.~\ref{LDE2}, which provides a demonstration of its use
in the instantaneous quench approximation in four-dimensional
field theory in flat space-time. The next obvious extension is to
couple the field to the scale factor of an expanding Universe.

%%%%%%%%%%%%%%%%%%%%%%%%%%%%%%%%%%%%%%%%%%%%%%%%%%%%%%%%%%%%%%%%%%%%%%%%
%%%%%%%%%%%%%%%%%%%%%%%%%%%%%%%%%%%%%%%%%%%%%%%%%%%%%%%%%%%%%%%%%%%%%%%%

\section*{Acknowledgements}
Thanks to Tim Evans, Ray Rivers, Nuno Antunes and Ed Copeland
for discussions, comments and criticisms. DJB was financially supported
by The Royal Commission for the Exhibition of 1851.

%%%%%%%%%%%%%%%%%%%%%%%%%%%%%%%%%%%%%%%%%%%%%%%%%%%%%%%%%%%%%%%%%%%%%%%%
%%%%%%%%%%%%%%%%%%%%%%%%%%%%%%%%%%%%%%%%%%%%%%%%%%%%%%%%%%%%%%%%%%%%%%%%

\section*{Appendix}
It is here demonstrated how to solve for the propagator
$iD(t_1,t_2)$ in quantum mechanics. We shall need to impose constraints
due to the
commutation relations and the KMS boundary condition, but we begin
by decomposing the propagator as
    \be
    i D(t_1,t_2)=\theta_c(t_1-t_2)iD^>(t_1,t_2)
    +\theta_c(t_2-t_1)iD^<(t_1,t_2)
    \ee
where $\theta_c(t-t')=\int_{t_0,c}^{t}{\rm d}t''\delta_c(t',t'')$.
Since $K(t)D(t,t')=\delta_c(t,t')$, it is straightforward to
demonstrate that
    \be
    K(t)D^{>(<)}(t,t')=0.
    \ee
We shall construct $D^{>(<)}$ from homogeneous solutions to the
quadratic operator $K$, i.e. functions which satisfy
$K(t)U^{\pm}(t)=0$. For $t<0$, these have the solution
$U^{\pm}(t)=\exp\{\pm i\omega_i t\}$. Thus, the most general form
for $D^{>(<)}$ is
    \be
iD^{>(<)}(t_1,t_2)=a^{>(<)}U^{+}(t_1)U^{-}(t_2)+b^{>(<)}U^{-}(t_1)U^{+}(t_2).
    \ee
Other possible combinations of $U^{\pm}$ can be ruled
out on imposing time translation invariance at early times. The
parameters $a^{>(<)}$ and $b^{>(<)}$ are to be determined. To do
this we begin by imposing the particle equal time commutation
relation
    \be
    [\hat{q},\hat{p}]=i.
    \ee
We make the free field identification $\langle T_c
\hat{q}(t_1)\hat{q}(t_2)\rangle=iD(t_1,t_2)$ and further that
$\hat{p}=\dot{\hat{q}}$. This leaves
    \be
\partial_{t_2}\left[iD^{>}(t_1,t_2)-iD^{<}(t_1,t_2)\right]_{t_1=t_2}=i
    \ee
which constrains the free parameters as follows
    \be
    a^{<}-a^{>}+b^{>}-b^{<}=\frac{1}{\omega_i}.
    \label{eq:1}
    \ee
A further symmetry requirement at equal time is that
    \be
    iD^{>}(t,t)=iD^{<}(t,t)
    \ee
which translates to
    \be
    a^>+b^>=a^<+b^<.
    \label{eq:2}
    \ee
Finally we impose the KMS boundary condition
    \be
    iD^<(t_0,t)=iD^>(t_0-i\beta,t)
    \ee
or
    \bea
    a^<&=&\exp\{\omega_i\beta\}a^>\label{eq:3}\\
    b^<&=&\exp\{-\omega_i\beta\}b^>\label{eq:4}.
    \eea
Eqs.~(\ref{eq:1}), (\ref{eq:2}), (\ref{eq:3}) and
(\ref{eq:4}) constitute 4 constraints on our 4 parameters. The set
of equations is easily solved yielding
    \bea
    a^>=b^<=\frac{1}{2\omega_i}\frac{1}{\exp\{\omega_i\beta\}-1}\\
a^<=b^>=\frac{1}{2\omega_i}\frac{\exp\{\omega_i\beta\}}{\exp\{\omega_i\beta\}-1}.
    \eea
We now have a general solution for the propagator at finite
temperature. Taking the zero temperature limit we have
    \bea
    iD^{>}(t_1,t_2)=\frac{1}{2\omega_i}U^{-}(t_1)U^{+}(t_2)\\
    iD^{<}(t_1,t_2)=\frac{1}{2\omega_i}U^{+}(t_1)U^{-}(t_2).
    \eea

The field theory case is much the same, with mode functions satisfying
\be
{K}_{\rm p}(t_1)D_{\rm p}(t_1,t_2)=\delta_c(t_1,t_2)
\ee
and
\be
\langle T_c\hat{\phi}_{\rm -p}(t_1)\hat{\phi}_{\rm
p}(t_2)\rangle=ViD_{\rm p}(t_1,t_2).
\ee
The solution for an initial state described by a temperature $1/\beta$
is
given in the main text.

%%%%%%%%%%%%%%%%%%%%%%%%%%%%%%%%%%%%%%%%%%%%%%%%%%%%%%%%%%%%%%%%%%%%%%%%%

%%%%%%%%%%%%%%%%%%%%%%%%%%%%%%%%%%%%%%%%%%%%%%%%%%%%%%%%%%%%%%%%%%%%%%%%%


\begin{thebibliography}{}

    \bibitem{GP}A.~H.~Guth and S.-Y.~Pi,
            Phys.~Rev. {\bf D 32}, 1899 (1985).

    \bibitem{COOP} F.~Cooper, S.-Y.~Pi and P.~N.~Stanicoff,
            Phys.~Rev. {\bf D 34}, 3831 (1986).

    \bibitem{ED} G.~J.~Cheetham and E.~J.~Copeland,
            Phys.~Rev. {\bf D 53}, 4125 (1996).

    \bibitem{HFJ} H.~F.~Jones, P.~Parkin and D.~Winder,
            Phys.~Rev. {\bf D 63}, 125013 (2001).

    \bibitem{BOY} D.~Boyanovsky, D.-S.~Lee and A.~Singh,
            Phys.~Rev. {\bf D 48}, 800 (1993).

    \bibitem{SCHW} J.~Schwinger,
            J. Math. Phys. 2, 407 (1961).

    \bibitem{KELD}L.~V.~Keldish,
            Sov.~Phys.~JETP 20, 1018 (1965).

\bibitem{LeBellac} M.~Le~Bellac,
            {\sl Thermal Field Theory},
            Cambridge University Press, 1996.

   \bibitem{Landsman} N.~P.~Landsman and Ch.~G.~van~Weert,
            Phys.~Rep. {\bf 145}, 141 (1987).

    \bibitem{PERN} S~.~A.~Pernice and G.~Oleaga,
            Phys. Rev. {\bf D 57}, 1144 (1998).

    \bibitem{Yukalov} V.~I.~Yukalov and E.~P.~Yukalova,
            Ann.~Phys.~(NY), {\bf 277}, 219 (1999).

    \bibitem{DJQM} I.~R.~C.~Buckley, A.~Duncan and H.~F.~Jones
            Phys.~Rev. {\bf D47}  2554 (1993).
            % LDE in QM

     \bibitem{DJAHO} A.~Duncan and H.~F.~Jones,
            Phys.~Rev. {\bf D47}  2560 (1993).
            %LDE in QFT

    \bibitem{ME} D.~J.~Bedingham and T.~S.~Evans,
            Phys.~Rev. {\bf D 64}, 105018 (2001).

    \bibitem{VCE} J.~O.~Akeyo and H.~F.~Jones
            Phys.~Rev. {\bf D47} 1668-1671 (1993).

   \bibitem{PJ} H.~F.~Jones and P.~Parkin
            Nucl.~Phys. {\bf B594 }  518 (2001).

   \bibitem{Berges} J.~Berges (private communication)

   \bibitem{gleiser}
        M.~Gleiser and R.~O.~Ramos,
        Phys. Rev. {\bf D50}, 2441 (1994).


   \bibitem{Boya}
        D.~Boyanovsky, H.~J.~de~Vega, R.~Holman,
        D.~S-Lee and A.~Singh, Phys. Rev. {\bf D51}, 4419 (1995).


   \bibitem{berges2}
        J.~Berges and J.~Cox, Phys. Lett. {\bf B517}, 369 (2001).

    \end{thebibliography}
\end{document}